\documentclass[12pt]{iopart}

\usepackage[dvips]{graphicx}
\usepackage{textcomp}
\usepackage{units}
\usepackage{url}
\usepackage[latin1]{inputenc} % to allow German characters as {\"a} to be typeset correctly

\newcommand{\CFS}{Co$_2$FeSi}

\def\vek#1{\mathbf{#1}}

\begin{document}

\title
{Ion beam induced modification of exchange interaction and
spin-orbit coupling in the Co$_2$FeSi Heusler compound}

\author{J. Hamrle, S. Blomeier, O. Gaier, B. Hillebrands}
\address{Fachbereich Physik and Forschungsschwerpunkt MINAS,
Technische Universit\"at Kaiserslautern,
Erwin-Schr\"odinger-Stra\ss e 56, D-67663 Kaiserslautern, Germany}

\author{H. Schneider, G. Jakob}
\address{Institut f\"ur Physik, Johannes
Gutenberg-Universit\"at Mainz, Staudinger Weg 7, D-55128 Mainz,
Germany}

\author{B. Reuscher, A. Brodyanski, M. Kopnarski}
\address{Institut f\"ur Oberfl\"achen- und Schichtanalytik,
Technische Universit\"at Kaiserslautern,
Erwin-Schr\"odinger-Stra\ss e 56, D-67663 Kaiserslautern, Germany}

\author{K. Postava}
\address{Department of Physics, Technical University of
Ostrava, 17. listopadu 15, 708 33, Ostrava-Poruba, Czech Republic}

\author{C. Felser}
\address{Institute of Inorganic and Analytical Chemistry,
Johannes Gutenberg-Universit\"at Mainz, Staudingerweg 9, D-55128
Mainz, Germany}

\date{\today}

\begin{abstract}

A Co$_2$FeSi (CFS) film with L2$_1$ structure was irradiated with
different fluences of \unit[30]{keV} Ga$^+$ ions. Structural
modifications were subsequently studied using the longitudinal
(LMOKE) and quadratic (QMOKE) magneto-optical Kerr effect. Both
the coercivity and the LMOKE amplitude were found to show a
similar behavior upon irradiation: they are nearly constant up to
ion fluences of \unit[$\approx6\times10^{15}$]{ion/cm$^2$}, while
they decrease with further increasing fluences and finally vanish
at a fluence of \unit[$\approx9\times10^{16}$]{ion/cm$^2$}, when
the sample becomes paramagnetic. However, contrary to this
behavior, the QMOKE signal nearly vanishes even for the smallest
applied fluence of \unit[$3\times10^{14}$]{ion/cm$^2$}. We
attribute this reduction of the QMOKE signal to an
irradiation-induced degeneration of second or higher order
spin-orbit coupling, which already happens at small fluences of
\unit[30]{keV} Ga$^+$ ions. On the other hand, the reduction of
coercivity and LMOKE signal with high ion fluences is probably
caused by a reduction of the exchange interaction within the film
material.

\end{abstract}

\maketitle

\section{Introduction}

Studies of ferromagnetic half-metals are mainly driven by their
possible applications for spintronic devices as a potential source
of a 100\% spin polarized current. Heusler alloys are promising
candidates for these applications due to their high Curie
temperature and expected half-metallicity even for partially
disordered systems.

Recently, the L2$_1$ ordered \CFS\ (CFS) Heusler compound
attracted a lot of attention from both experimental and
theoretical points of
view.\cite{wur06apl,ino06,kal06,kan06,has05,wur05,nic79} Up to
now, CFS exhibits the highest observed magnetic moment
(\unit[5.97]{$\mu_B$}) per formula unit (at \unit[5]{K},
corresponding to an average value of \unit[1.49]{$\mu_B$} per
atom) and the highest Curie temperature (\unit[1100]{K}) among the
so-called Heusler compounds and half-metallic
ferromagnets.\cite{wur06apl} CFS grows fully epitaxial on MgO(001)
forming a L2$_1$ structure.\cite{wur06,schneider06} A very high
tunnelling magnetoresistance of 41\% at RT and 60\% at 5\,K were
reported for CFS/Al-oxide/Co$_{75}$Fe$_{25}$ samples,\cite{ino06}
whose crystalline structure was unclear, however. Moreover, time
resolved photoemission studies resolving the spin polarization at
the Fermi level indicated a rather low value of only 12\% for CFS
films grown on MgO(001) substrates.\cite{schneider06} Further
studies revealed that the observed reduction of the spin
polarization cannot only be attributed to surface effects, as it
also takes place up to about \unit[5]{nm} away from the surface.
\cite{wus06} Moreover, recent studies showed that CFS films
exhibit a huge quadratic magneto-optical Kerr effect (QMOKE) with
amplitudes of up to \unit[30]{mdeg}, which is the largest QMOKE
observed so far. This effect is a fingerprint of an unusually
large spin-orbit coupling of second or higher order.
\cite{hamrle-CFS-QMOKE}

In order to understand the origin of these intriguing
magneto-optical properties, it is advantageous to have a method at
hand which can be used to tune them in a controlled way. It has
been shown that irradiation with keV ions is an efficient method
to modify the properties of many magnetic thin film systems, such
as CoPt multilayers with perpendicular
anisotropy,\cite{Chappert1998, Devolder1999, Hyndman2001,
Warin2001} exchange bias bilayers,\cite{Mougin2001,
Fassbender2003, Fassbender2004} or interlayer exchange coupled
trilayers.\cite{Demokritov2003, Demidov2004, Blomeier2005}
Moreover, it was demonstrated that such modifications can be
performed with high spatial resolution, making ion irradiation an
attractive patterning tool in view of technological
applications.\cite{Chappert1998, Devolder1999, Fassbender2003,
Fassbender2004, Blomeier2005, Blomeier2006} In this work we
demonstrate that the magnetic and, in particular, the
magneto-optical properties of CFS Heusler alloys can also be
modified by means of keV ion irradiation.

\section{Sample properties and preparation}
\label{s:sample}

We have investigated a CFS film of thicknesses of \unit[11]{nm}
prepared by RF magnetron sputtering and deposited directly onto
MgO(001). The film was covered by a \unit[4]{nm} thick Al
protective layer. As mentioned in the introduction, these films
grow in the L2$_1$ ordered structure. A detailed description of
the sample preparation process as well as structural properties of
CFS films can be found in Ref.\cite{schneider06}. On the sample, 9
different areas of \unit[$\approx1$]{mm$^2$} size were defined,
with eight of them being irradiated with different fluences of
\unit[30]{keV} Ga$^+$ ions varying from
\unit[$3\times10^{14}$]{ion/cm$^2$} to
\unit[$9\times10^{16}$]{ion/cm$^2$}. The ninth area was left
unirradiated for reference purposes.

\section{CFS irradiated by G\lowercase{a}$^+$ ions}
\label{s:III}

\begin{figure}
\begin{center}
\includegraphics[width=0.6\textwidth]{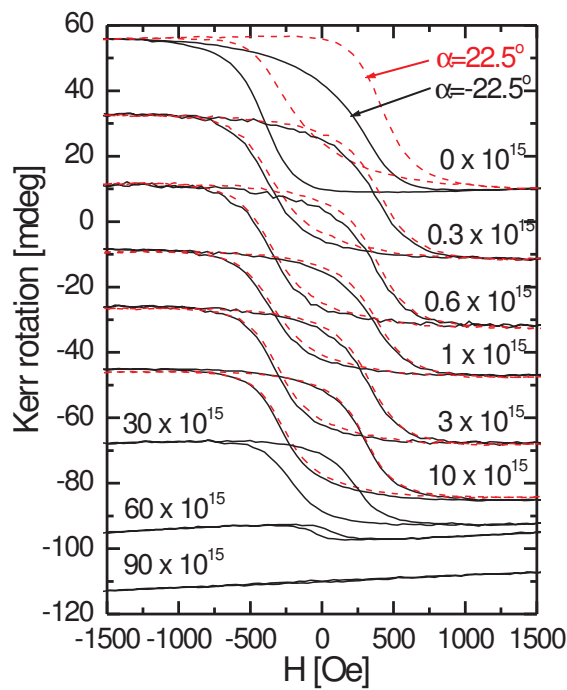}
\end{center}
\caption{%
\label{f:irradloops}%
(color online) MOKE hysteresis loops recorded from different areas
of the CFS(\unit[11]{nm}) sample that were irradiated with
different fluences of \unit[30]{keV} Ga$^+$ ions. The sample
orientation $\alpha$ is equal to $22.5^\circ$ (dashed red line) or
$-22.5^\circ$ (full black line), respectively.}
\end{figure}

Magneto-optical Kerr effect (MOKE) hysteresis loops recorded from
sample areas with different irradiation doses are shown in
Fig.~\ref{f:irradloops}. The loops are measured at the incidence
angle $\varphi=45^\circ$ using s-polarized light at a wavelength
of \unit[$\lambda=670$]{nm}. The measured quantity is the Kerr
rotation. The loops are presented for a sample orientation angle
$\alpha\pm22.5^\circ$, which is the angle between the [100] CFS
direction and the plane of incidence of light. The loops in
Fig.~\ref{f:irradloops} exhibit several interesting features: (i)
Both the coercivity $H_c$ and the longitudinal MOKE (LMOKE) loop
amplitude are decreasing as the ion fluence increases. The sample
area which was irradiated with the highest fluence of
\unit[$9\times 10^{16}$]{ions/cm$^2$} exhibits paramagnetic
properties. (ii) The loop recorded from the nonirradiated area
shows a large loop asymmetry. This asymmetry originates from a
superimposed QMOKE contribution,
\cite{ost98,pos97,pos02,cow97,pos03,mew04} which results in an
even contribution to the measured MOKE loop. All loops recorded
from irradiated areas exhibit a much smaller asymmetry, i.e., a
much smaller QMOKE contribution. (iii) The loop asymmetry (i.e.,
the indication of QMOKE contribution) is reverted when $\alpha$ is
changed from $22.5^\circ$ to $-22.5^\circ$ reflecting the
characteristic fourfold symmetry of the QMOKE originating from a
cubic crystal. \cite{pos02,cow97,pos03,mew04,vis86}

The dependence of $H_c$, as well as the LMOKE and QMOKE amplitudes
on the applied ion fluence is given in Fig.~\ref{f:irrad}. $H_c$
(Fig.~\ref{f:irrad}(a)) and the LMOKE amplitude
(Fig.~\ref{f:irrad}(b)) show a similar behavior; they are
decreasing very slowly up to an ion fluence of
$\approx$\unit[$6\times 10^{15}$]{ions/cm$^2$}. For higher
fluences, both $H_c$ and LMOKE decrease faster and finally vanish
at a fluence of \unit[$9\times 10^{16}$]{ions/cm$^2$} when the
sample becomes paramagnetic. This behavior shows that the sample
retains ferromagnetic properties up to high irradiation fluences,
which is an indication that these properties are resistant to a
certain degree of atomic disorder in the CFS structure.

\section{QMOKE amplitude}

\begin{figure}
\begin{center}
\includegraphics[width=1\textwidth]{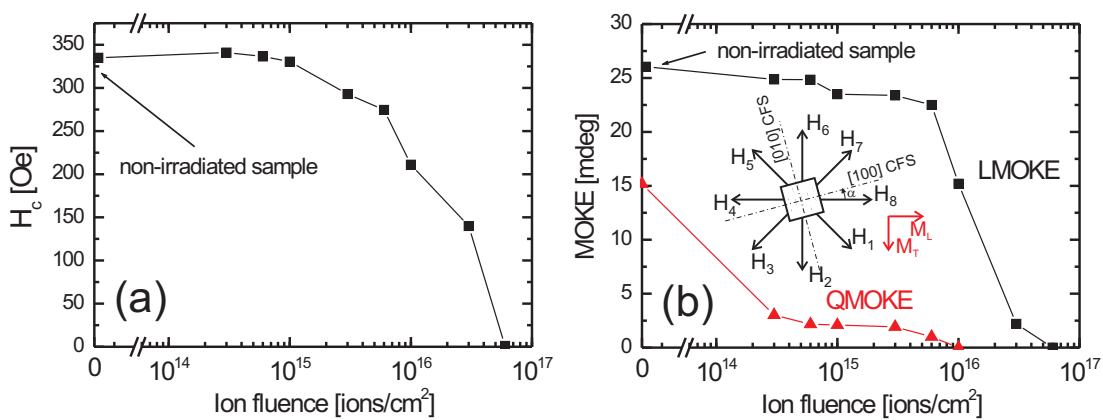}
\end{center}
\caption{%
\label{f:irrad}%
(color online) Dependence of (a) $H_c$ and (b) the amplitudes of
the LMOKE and QMOKE of the CFS(\unit[11]{nm}) sample on the
applied ion fluence. The inset in (b) sketches the in-plane
magnetization directions used to determine the QMOKE amplitude in
saturation. \cite{pos02,pos03}. See text for details.}
\end{figure}

\begin{figure}
\begin{center}
\includegraphics[width=0.4\textwidth]{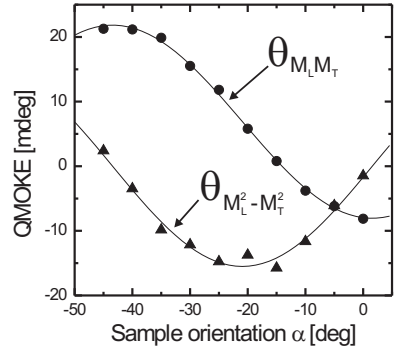}
\end{center}
\caption{%
\label{f:QMOKEalpha}%
Dependence of the QMOKE Kerr rotations $\theta_{M_LM_T}$ and
$\theta_{M_L^2-M_T^2}$ on the sample orientation $\alpha$ measured
on the non-irradiated sample area. Full lines are fits to
Eq.~(\ref{eq:MLMT}). }
\end{figure}

As discussed in point (ii), Sect.~\ref{s:III}, there is a large
suppression of the QMOKE signal even for small applied doses. In
order to determine the amplitude of the pure QMOKE effect
quantitatively, the 8-directional procedure described in
\cite{pos02, hamrle-CFS-QMOKE} is used. Within this procedure, the
in-plane magnetic field is subsequently applied in directions from
$\vek{H}_1$ to $\vek{H}_8$, defined in the inset of
Fig.~\ref{f:irrad}(b). In order to almost suppress the LMOKE
signal, a nearly normal angle of incidence of light is used
(incidence angle is about $0.5^\circ$). The QMOKE consists of two
contributions, being proportional to $M_LM_T$ and $M_L^2-M_T^2$
terms, where $M_L$ and $M_T$ is the longitudinal and transverse
direction, respectively, as defined in the inset of
Fig.~\ref{f:irrad}(b). The contribution to the QMOKE Kerr rotation
proportional to $M_LM_T$ is determined from
$\theta_{M_LM_T}=[\theta(\vek{H}_1)+\theta(\vek{H}_5)-\theta(\vek{H}_3)-\theta(\vek{H}_7)]/4$
whereas the QMOKE Kerr rotation proportional to $M_L^2-M_T^2$
writes
$\theta_{M_L^2-M_T^2}=[\theta(\vek{H}_8)+\theta(\vek{H}_4)-\theta(\vek{H}_2)-\theta(\vek{H}_6)]/4$,
where $\theta(\vek{H}_i)$ is the Kerr rotation in a given
$\vek{H}_i$ direction. The QMOKE Kerr rotations $\theta_{M_LM_T}$
and $\theta_{M_L^2-M_T^2}$ are measured for several sample
orientation $\alpha$, which is the angle between the [100] CFS
axis and the $\vek{H_8}$ direction (see inset in
Fig.\ref{f:irrad}(b)). The dependence of $\theta_{M_LM_T}$ and
$\theta_{M_L^2-M_T^2}$ on $\alpha$ for the non-irradiated sample
area are presented in Fig.~\ref{f:QMOKEalpha}. In agreement with
theory, \cite{pos02, pos03, hamrle-CFS-QMOKE} it is evident that
$\theta_{M_LM_T}$ and $\theta_{M_L^2-M_T^2}$ can be described as
\begin{equation}
\begin{array}{rl}
 \label{eq:MLMT}
\displaystyle \theta_{M_LM_T}\!\!& \displaystyle =Q_A \cos
4\alpha+Q_B
\\
\displaystyle \theta_{M_L^2-M_T^2}\!\!&\displaystyle =Q_A \sin
4\alpha\quad .
\end{array}
\end{equation}
The four-fold symmetry of Eq.~(\ref{eq:MLMT}) reflects the cubic
symmetry of the CFS crystal. The quantity $Q_A$ in
Eq.~(\ref{eq:MLMT}) represents the amplitude of variation of the
QMOKE signal with $\alpha$ whereas $Q_B$ is a constant
contribution to the QMOKE signal independent on sample
orientation.

The dependence of the QMOKE amplitude $Q_A$ on the applied ion
fluence is presented in Fig.~\ref{f:irrad}(b). The QMOKE amplitude
of the non-irradiated area is equal to \unit[15]{mdeg} whereas it
decreases rapidly to only \unit[3]{mdeg} for the smallest ion
fluence of \unit[$3\times 10^{14}$]{ions/cm$^2$}. The constant
term $Q_B$ (not presented in Fig.~\ref{f:irrad}(b)) also provides
a similar drop as $Q_A$ with applied ion fluence. It was checked
that this rapid decrease of the QMOKE amplitude with the applied
ion fluence also appears for the corresponding Kerr ellipticity.
Therefore, we conclude that the drop of the QMOKE signal upon
irradiation is related to a change of the electronic structure of
CFS. Contrary to LMOKE, the QMOKE signal appears to be very
sensitive to ion irradiation and, therefore, to atomic disorder of
the CFS structure.

\section{Discussion}

\begin{figure}
\includegraphics[width=0.98\textwidth]{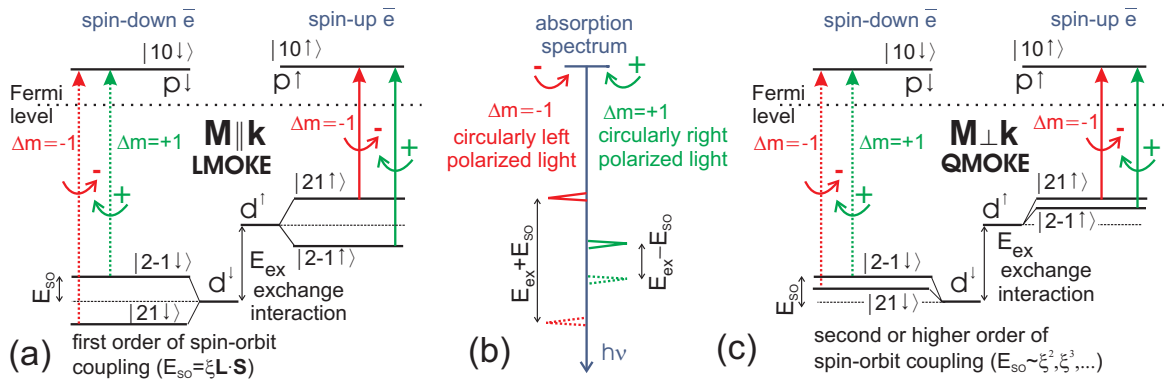}
\caption{%
\label{f:micro}%
(color online) Simplified sketch of the electronic structure for
one point in $\vek{k}$-space for (a) $\vek{M}\parallel\vek{k}$ and
(c) $\vek{M}\perp\vek{k}$, giving rise to LMOKE and QMOKE
respectively. (b) illustrates absorption spectra of the dipolar
(optical) transitions presented in (a). The presence of a Kerr
effect requires that the absorption spectra must be different for
left ($\Delta m=-1$) and right ($\Delta m=1$) circularly polarized
light, and therefore both exchange and spin-orbit interaction must
be present. For more details see Ref. \protect\cite{ost98,bru96}.}
\end{figure}

It is known that QMOKE and LMOKE effects are related to the
electronic structure of a given sample in different
ways.\cite{ost98,bru96,hul32} Fig.~\ref{f:micro}(a) presents a
simplified sketch of electronic structure of a ferromagnetic
material for one point of $\vek{k}$-space.\cite{bru96} Here, we
limit ourselves only to $d\rightarrow p$ transitions, we assume no
exchange between $p$-states, and we show only dipolar (i.e.,
optical) transitions to $p$ states $|10\rangle$. In our simplified
sketch, the exchange interaction $E_\mathrm{ex}$ splits only the
$d$ states into $d^\uparrow$ and $d^\downarrow$ for up- and
down-electron levels, respectively. Furthermore, both $d^\uparrow$
and $d^\downarrow$ states are split by spin-orbit (SO) coupling
$E_\mathrm{SO}$ accordingly to the magnetic quantum number $m$ of
electrons in $d$ states. Depending on the difference $\Delta m$ of
the magnetic quantum numbers between the final (here $p$) and
initial (here $d$) states, circularly left ($\Delta m=-1$) or
circularly right ($\Delta m=1$) polarized photons can be absorbed
by the electronic structure. The absorbtion spectra for both
polarizations are sketched in Fig.~\ref{f:micro}(b). When both
exchange and SO coupling are present, the absorption spectra for
the left and right polarized light are different. In such a case,
a non-zero Kerr effect arises. \cite{bru96} However, when either
exchange or SO coupling are not present (i.e., $E_\mathrm{ex}=0$
\emph{or} $E_\mathrm{SO}=0$), then the absorption spectra are
identical, providing zero Kerr effect.

From a microscopic point of view, the LMOKE originates from a
component of $\vek{M}$ parallel to $\vek{k}$ whereas the QMOKE
originates from a component of $\vek{M}$ perpendicular to
$\vek{k}$, where $\vek{k}$ denotes the vector of propagation of
light in matter.\cite{ost98} In the following we discuss changes
in the electronic structure for $\vek{M}\parallel\vek{k}$ (related
to LMOKE) and $\vek{M}\perp\vek{k}$ (related to QMOKE), as
sketched in Fig.~\ref{f:micro}(a,c).

In the case of $\vek{M}\parallel\vek{k}$, the SO coupling is
proportional to the spin-orbit coupling parameter $\xi$,
$E_\mathrm{SO}=\xi \vek{L} \cdot \vek{S}$ (Fig.~\ref{f:micro}(a)).
On the other hand, in the case of $\vek{M}\perp\vek{k}$, the first
order of SO coupling in $\xi$ is zero, $\xi
\vek{L}\cdot\vek{S}=0$. Therefore only spin-orbit effects of
second or higher order are able to remove the degeneracy of the
initial or final state. \cite{ost98} Usually, higher orders of SO
coupling are much smaller than the first order,
$\xi\vek{L}\cdot\vek{S}$, leading to a much smaller QMOKE
amplitude in comparison with the corresponding LMOKE amplitude in
most materials.

According to the previous discussion, the suppression of the LMOKE
amplitude by means of large applied ion fluences (from
\unit[$\approx6\times10^{15}$]{ion/cm$^2$}) may be understood
either in terms of a reduction of exchange interaction or a
reduction of the first order of SO coupling. As the LMOKE exhibits
a behavior similar to that of $H_c$, we may conclude that it is
the exchange interaction in CFS which is reduced by large fluences
of Ga$^+$ ions. On the other hand, the first order of SO coupling
seems to persist at such high fluences. If SO coupling disappears
before exchange, a decreasing LMOKE amplitude should be observed
with increasing fluence but the magnetic properties of the sample
should not vary drastically.

In the case of non-irradiated CFS films, an unusually large QMOKE
amplitude was measured, indicating a presence of a large SO
coupling of second or higher order. Furthermore, small ion
fluences solely reduce the QMOKE amplitude. Therefore, we may
conclude that second or higher order contributions to SO coupling
are very sensitive to small fluences of Ga$^+$ ion irradiation in
CFS. However, a deeper microscopic reason for this effect is not
yet clear.

\section{Summary}

The effect of \unit[30]{keV} Ga$^+$ ion irradiation on epitaxial
\CFS (CFS) film having L2$_1$ structure and deposited onto
MgO(001) was studied by means of the longitudinal (LMOKE) and
quadratic (QMOKE) magneto-optical Kerr effect. Both the coercivity
and the LMOKE exhibit a similar behavior on the applied ion
fluence: they are nearly constant up to an ion fluence of
\unit[$\approx6\times10^{15}$]{ion/cm$^2$}, after which they
decrease and finally vanish at a fluence of
\unit[$\approx9\times10^{16}$]{ion/cm$^2$}, when the sample
becomes paramagnetic. The fluence dependence of the QMOKE signal
is very different: Its amplitude drops from \unit[15]{mdeg} for a
non-irradiated sample region to \unit[3]{mdeg} for the smallest
applied fluence of \unit[$3\times10^{14}$]{ion/cm$^2$}. The
observed reduction of the QMOKE signal is attributed to an
irradiation-induced degeneration of second or higher order
contributions to the spin-orbit coupling, which already happens at
small ion fluences. On the other hand, the reduction of coercivity
and LMOKE at high ion fluences can be attributed to a reduction of
exchange interaction within the irradiated areas.

\section{Acknowledgment}

The project was financially supported by the Research Unit 559
\emph{"New materials with high spin polarization"} funded by the
Deutsche Forschungsgemeinschaft, and by the Stiftung
Rheinland-Pfalz f\"ur Innovation. Partial support from the Grant
Agency of the Czech Republic (202/06/0531) and from the NEDO
project of the Japanese government is acknowledged. We would like
to thank T. Mewes for stimulating discussions.

\newpage
\section{References}

%\bibliography{heubibII}

\begin{thebibliography}{10}

\bibitem{wur06apl}
Wurmehl S, Fecher GH, Kandpal HC, Ksenofontov V, Felser C, Lin HJ
2006 {\em
  Appl. Phys. Lett.} {\bf 88} 032503.

\bibitem{ino06}
Inomata K, Okamura S, Miyazaki A, Kikuchi M, Tezuka N, Wojcik M,
Jedryka E 2006
  {\em J. Phys. D: Appl. Phys.} {\bf 39} 816.

\bibitem{kal06}
Kallmayer M, Elmers HJ, Balke B, Wurmehl S, Emmerling F, Fecher
GH, Felser C
  2006 {\em J. Phys. D: Appl. Phys.} {\bf 39} 786.

\bibitem{kan06}
Kandpal HC, Fecher GH, Felser C, Sch{\"o}nhense G 2006 {\em Phys. Rev.
B} {\bf 73}
  094422.

\bibitem{has05}
Hashimoto M, Herfort J, Sch{\"o}nherr HP, Ploog KH 2006 {\em Appl.
Phys. Lett.}
  {\bf 87} 102506.

\bibitem{wur05}
Wurmehl S, Fecher GH, Kandpal HC, Ksenofontov V, Felser C, Lin HJ,
Morais J
  2005 {\em Phys. Rev. B} {\bf 72} 184434.

\bibitem{nic79}
Niculescu V, Budnick JI, Hines WA, Raj K, Pickart S, Skalski S
1979 {\em Phys.
  Rev. B} {\bf 19} 452.

\bibitem{wur06}
Wurmehl S, Fecher GH, Kroth K, Kronast F, D{\"u}rr HA, Takeda Y,
Saitoh Y,
  Kobayashi K, Lin HJ, Sch{\"o}nhense G, Felser C 2006 {\em J. Phys. D: Appl.
  Phys.} {\bf 39} 803.

\bibitem{schneider06}
Schneider H, Jakob G, Kallmayer M, Elmers HJ, Cinchetti M, Balke
B, Wurmehl S,
  Felser C, Aeschlimann M, Adrian H 2006 {\em Phys. Rev. B} {\bf 74}
  174426.

\bibitem{wus06}
W\"ustenberg JP, Cinchetti M, Albaneda MS, Bauer M, Aeschlimann M
2006 2006 {\it J.~Magn.~Magn.~Mater.} in press,
  arXiv:cond-mat/0606006.

\bibitem{hamrle-CFS-QMOKE}
Hamrle J, Blomeier S, Gaier O, Hillebrands B 2006
  arXiv:cond-mat/cond-mat/0609688,.
\newblock To be published

\bibitem{Chappert1998} Chappert C, Bernas H, Ferr\'{e} J,
Kottler V, Jamet J-P, Chen Y, Cambril E, Devolder T, Rousseaux F,
Mathet V, Launois H, 1998 {\it Science} {\bf 280} 1919.

\bibitem{Devolder1999} Devolder T, Chappert C, Chen Y, Cambril E,
Bernas H, Jamet J-P, Ferr\'{e} J, 1999 {\it Appl.~Phys.~Lett.}
{\bf 74} 3383.

\bibitem{Hyndman2001} Hyndman R, Warin P, Gierak J, Ferr\'{e} J,
Chapman JN, Jamet J-P, Mathet V, Chappert C, 2001 {\it
J.~Appl.~Phys.} {\bf 90} 3843.

\bibitem{Warin2001} Warin P, Hyndman R, Gierak J, Chapman JN,
Ferr\'{e} J, Jamet J-P, Mathet V, Chappert C, 2001 {\it
J.~Appl.~Phys.} {\bf 90} 3850.

\bibitem{Mougin2001} Mougin A, Mewes T, Jung M, Engel D, Ehresmann A,
Schmoranzer H, Fassbender J, Hillebrands B, 2001 {\it
Phys.~Rev.~B} {\bf 63} 060409(R).

\bibitem{Fassbender2003} Fassbender J, Poppe S, Mewes T, Juraszek J,
Hillebrands B, Barholz K-U, Mattheis R, Engel D, Jung M,
Schmoranzer H, Ehresmann A, 2003 {\it Appl.~Phys. A} {\bf 77} 51.

\bibitem{Fassbender2004} Fassbender J, Ravelosona D, Samson Y,
2004 {\it J.~Phys.~D} {\bf 37} R179.

\bibitem{Demokritov2003} Demokritov SO, Bayer C, Poppe S,
Rickart M, Fassbender J, Hillebrands B, Kholin DI, Kreines NM,
Liedke OM, 2003 {\it Phys.~Rev.~Lett.} {\bf 90} 097201.

\bibitem{Demidov2004} Demidov VE, Kholin DI, Demokritov SO,
Hillebrands B, Wegelin F, Marien J, 2004 {\it Appl.~Phys.~Lett.}
{\bf 84} 2853.

\bibitem{Blomeier2005} Blomeier S, Hillebrands B, Demidov VE,
Demokritov SO, Reuscher B, Brodyanski A, Kopnarski M, 2005 {\it
J.~Appl.~Phys.} {\bf 98} 093503.

\bibitem{Blomeier2006} Blomeier S, Candeloro P, Hillebrands B,
Reuscher B, Brodyanski A, Kopnarski M: 2006 {\it
J.~Magn.~Magn.~Mater.} in press.

\bibitem{ost98}
{Osgood III} RM, Bader SD, Clemens BM, White RL, Matsuyama H 1998
{\em J. Magn.
  Magn. Mater.} {\bf 182} 297.

\bibitem{pos97}
Postava K, Jaffres H, Schuhl A, {Nguyen Van Dau} F, Goiran M, Fert
AR 1997 {\em
  J. Magn. Magn. Mater.} {\bf 172} 199.

\bibitem{pos02}
Postava K, Hrabovsk\'y D, Pi\v{s}tora J, Fert AR,
Vi\v{s}\v{n}ovsk\'y {\v{S}},
  Yamaguchi T 2002 {\em J. Appl. Phys.} {\bf 91} 7293.

\bibitem{cow97}
Cowburn RP, Ferr\'{e} J, Jamet JP, Gray SJ, Bland JAC 1997 {\em
Phys. Rev. B}
  {\bf 55} 11593.

\bibitem{pos03}
Postava K, Pi\v{s}tora J, Yamaguchi T, Hlubina P.
\newblock Polarized light in structures with magnetic ordering.
\newblock In: Pluta M, Szyjer M, Powichrowska E, editors. Lightmetry 2002:
  {M}etrology and {T}esting {T}echniques {U}sing {L}ight vol. 5064 of Proc. of
  SPIE Bellingham, Wash.: SPIE; 2003. p. 182--190.

\bibitem{mew04}
Mewes T, Nembach H, Rickart M, Hillebrands B 2004 {\em J. Appl.
Phys.} {\bf 95}
  5324.

\bibitem{vis86}
Vi\v{s}\v{n}ovsk\'y {\v{S} } 1986 {\em Czech. J. Phys. B} {\bf 36}
625.

\bibitem{bru96}
Bruno P, Suzuki Y, Chappert C 1996 {\em Phys. Rev. B} {\bf 53}
9214.

\bibitem{hul32}
Hulme HR 1932 {\em Proc. R. Soc. London} {\bf A135} 237.

\end{thebibliography}
%\bibliographystyle{vancouver-jaro}

\clearpage

%%%%%%%%%%%%%%%%%%%%%%%%%%%%%%%%%%%%%%%%%%%%%%%%%%%%%%%%%%%%%%%%%

%%%%%%%%%%%%%%%%%%%%%%%%%%%%%%%%%%%%%%%%%%%%%%%%%%%%%%%%%%%%%%%%%

%%%%%%%%%%%%%%%%%%%%%%%%%%%%%%%%%%%%%%%%%%%%%%%%%%%%%%%%%%%%%%%

\end{document}